\documentclass[twoside]{article}

\usepackage{PRIMEarxiv}

\usepackage[utf8]{inputenc} 
\usepackage[T1]{fontenc}    
\usepackage{hyperref}       
\usepackage{url}            
\usepackage{booktabs}       
\usepackage{amsfonts}       
\usepackage{nicefrac}       
\usepackage{microtype}      
\usepackage{lipsum}
\usepackage{natbib}
\usepackage{graphicx}       
\graphicspath{{media/}}     

\title{Managing Controlled Unclassified Information in Research Institutions
}

\author{
Baijian Yang, Preston Smith, Huyunting Huang\\
Purdue University \\
West Lafayette, IN  \\
\texttt{\{byang, psmith, huan1182\}purdue.edu} \\
\And
 Carolyn Ellis \\
University of California San Diego  \\
La Jolla, CA 92093\\
\texttt{c1ellis@ucsd.edu} \\
}

\begin{document}
\maketitle

\begin{abstract}

In order to operate in a regulated world, researchers need to ensure compliance with  ever-evolving landscape of information security regulations and best practices. 
This work explains the concept of Controlled Unclassified Information (CUI) and the challenges it brings to the research institutions.   
Survey from the user perceptions showed that most researchers and IT administrators lack a good understanding of CUI and how it is related to other regulations, such as  HIPAA, ITAR, GLBA, and FERPA. 
A managed research ecosystem is introduced in this work. The workflow of this efficient and cost effective framework is elaborated to demonstrate how controlled research data are processed to be compliant with one of the highest level of cybersecurity in a campus environment. 
Issues beyond the framework itself is also discussed. 
The framework serves as a reference model for other institutions to support CUI research. 
The awareness and training program developed from this work will be shared with other institutions to build a bigger CUI ecosystem.

\end{abstract}


\section{Introduction}

Research compliance is a fundamental part of the research life-cycle that needs to be addressed from inception, execution, and after a research project is completed. 
In order to operate in a regulated world, researchers will face the task of understanding the landscape of information security regulations and best practices. In the US, regulations like HIPAA \cite{herold2003practical}, ITAR \cite{blount2008itar}, and FERPA \cite{ramirez2009ferpa} are all in the landscape and must be addressed. Outside the US jurisdiction, other security laws are drafted and passed. A good example is  GDPR \cite{voigt2017eu}, which is so far the toughest law to product individual privacy. While the data protection may sound like an IT issue, the impact is much broader than it seems. In particular, both the researchers and campus IT are often at lost when it comes to the data protection needed to meet the requirements of different regulations. The Defense Acquisition Regulation Supplement (DFARS) gives clear direction with the issuance of Executive Order 13556 \cite{order201013556}, “Controlled Unclassified Information,” any federal information that requires safeguarding or dissemination controls pursuant to or consistent with applicable law, regulations or government-wide policies, is now defined as Controlled Unclassified Information (CUI).  Categories of CUI include, but are not limited to healthcare data subject to HIPAA,  student data subject to FERPA, and defense data subject to ITAR. Sponsors are increasingly requiring that such information held by non-federal entities be secured in line with NIST SP 800-171 \cite{ross2020enhanced}. Private sector sponsors often require assurances that their data or methods used in university research will be appropriately safeguarded, to criteria of the sponsor’s own design. However, the general Federal Acquisition Regulation (FAR) has not been decided and now federal and state agencies add their own requirements, which causes extra work for research administrators and complexity for researchers. 

In the international scope, ISO 27001 and 27002 \cite{song2014characteristics} are the most recognized standards that spell out the specifications for Information Security Management System (ISMS). Under ISO 27001, a system is well protected only when security controls are implemented to protect the following five categories of assets: \textit{data, hardware, software, people and network}. While ISO 27001 is a certification standard, ISO 27002 is a guideline that offers best practices to implement ISO 27001 \cite{disterer2013iso}. Although NIST SP 800-171 and ISO 27001 have differences in implementations, both standards cover the same area in information management. In the version published in 2015 \cite{ross2015protecting}, NIST has attempted to to map all key security controls to ISO 27001 in NIST SP 800-171 rev1. Although the mapping was not prefect and was dropped from its most recent revisions, there are many similarities between the two standards.  Donaldson (2015) also compared cybersecurity frameworks developed by ISC2, ISO, NIST and CCSCC  \cite{donaldson2015cybersecurity}. It was stated that major topics and categories are consistent across all frameworks, in spit of the differences on wording and content organizations. All four frameworks break the protection into functional areas; all employed the risk management mythology; all include security control; and all provide a mechanism to audit, evaluate and validate the security control. 
This suggests that if a guideline is developed to meet the NIST SP 800-171 requirement, it can be further revised to help an organization to meet the certification requirements of ISO 27001.

CUI is not an additional classification as outlined in the Classified National Security Information program. The required protections are defined in NIST publication 800-171.While the controls are designed to ensure confidentiality of the data set, it is up to the institution to design the remaining controls to ensure integrity and availability of the information systems hosting CUI. Similar federal programs provide frameworks such as the Risk Management Framework (RMF) \cite{ting2010information} in conjunction with NIST publications to provide an appropriate security model used to ensure confidentiality, integrity and availability. \emph{By design, there is not a recommended framework such as RMF to help institutions integrate the appropriate controls as they are outlined in NIST SP 800-171, as the publication is not considered a standalone framework.} It is currently up to each institution to create supplemental mappings to identity and fix gaps to ensure integrity and availability. 
In this paper, we present a scalable framework that is transparent to the stakeholders and is easily extended to other campus computing environment. 
The lightweight framework is based on NIST 800-171 and supplemented by other NIST publications.  It serves as a 
reference model for other institutions to support research CUI. The framework increases the overall security posture of the information systems controlled by the framework, while simplifying the implementation and monitoring processes.
The proposed framework, in its current stage, is a guideline to support CUI management for institutions in the US. Due to the similarity between  NIST 800-171 and ISO 27001, the proposed the CUI framework has the potential to be revised and applied to research institutions outside the US.
When establishing the framework, we use a staged approach and input from various contributors to ensure this remains an iterative approach in order to establish the strongest possible connection to quality and ensure the security controls reach their highest maturity level.
The framework provides a standard methodology and assessment scale. The goals of our design are as follows. 
\begin{itemize}
    \item Empower Campus IT With a Standard Campus Framework for Data Security by creating a campus framework for research cybersecurity based on NIST SP 800-171. 
    \item Improve Processes for Research Administration: Develop a single process for intake, and contracting, and facilitate easy mapping to CI resources for the sponsored programs office, human subjects office, and export control office.
    \item Educate Researchers on Regulations and Cybersecurity Practices by creating materials to train scientists on regulatory requirements set by funding agencies.
    \item Develop cyberinfrastructure Professionals through undergraduate student participation in creation of the framework. By pairing students with staff mentors and area Information Systems Security Managers (ISSMs), this work prepares them to become skilled cybersecurity professionals ready to contribute as they join the workforce.
\end{itemize}

The rest of the paper is organized as follows.  Background information and related work will be reviewed in Section 2, followed by a user perception study in Section 3. The CUI management workflow is then elaborated in details in Section 4. After discussions in Section 5, this paper concludes in Section 6.   

\section{Background Information}
Universities and research institutes traditionally have enjoyed wide academic freedom with respect to how they handled data, particularly in the realm of fundamental research, which may be publicly released. This is changing with data breaches becoming more frequent and common. Various U.S. government agencies have progressively tightened the security requirements for handling research data. Industrial partners, while not as regulated by statute, frequently require university IT practices to align with their own corporate policies. 

With the guidelines from DRFARS, research projects that may fall into different categories. For examples, a research project that may take input data from radar, satellites, reconnaissance aircraft and ships is subjected to CUI for export controlled research. A research project that uses quantitative, qualitative, and mixed methods designs to maximize existing support structures and enhancing workplace policies to support working caregivers falls into the category of CUI for health information. A research project to conduct experiments in plant phenomics in a state-of-the-art phenotyping facility is under the CUI Agriculture category of the intelligence group. A basic science research project that has applications in aviation and power generation industries will find themselves need to protect the research data in the interest of CUI category of General Proprietary Business Information.  

As of April 2020, CUI categories are organized into the following 20 indexing groups: critical infrastructure, defense, export control, financial, immigration, intelligence, international agreements, law enforcement, legal, natural and cultural resources, North Atlantic Treaty Organization (NATO), nuclear, patent, privacy, procurement and acquisition, proprietary business information, provisional, statistical, tax, and transportation. Each group has one or multiple categories. The total number of categories is as many as 125, which could become a barrier for researchers to properly conduct their research. 

To support open science, many research institutions created their own campus cyberinfrastructure. Consisting of high-performance computing, project data storage, and archival storage, the cyberinfrastructure is designed to serve the broadest set of science communities possible and is a partner in national and worldwide communities such as the Open Science Grid, XSEDE, and the LHC Computing Grid. In addition to the cyberinfrastructure serving open science, some universities have begun their journeys to build an ecosystem dedicated to broadly supporting regulated research. 
For instance, 
ResShield and ResVault \cite{UF2017} were built to meet FISMA \cite{nowell2007regulatory} compliance at the moderate impact level, which implies CUI compliance as specified in 800-171.
A framework called Research Ecosystem for Encumbered Data (REED) to support CUI compliance \cite{reed2017} was also built to support ITAR research in community cluster as well as Amazon AWS GovCloud. 

\begin{figure}[tb]
    \centering
    \includegraphics[width=.90\textwidth]{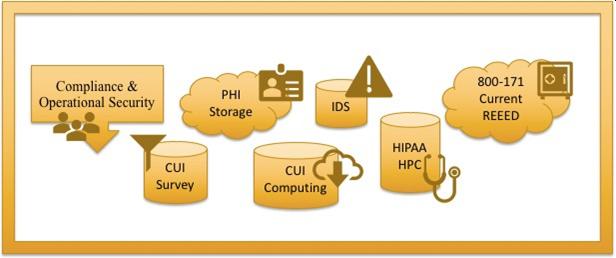}
    \caption{The Components of a CUI Ecosystem}
    \label{fig:eco}
\end{figure}

Both systems aims to build a CUI ecosystem, as illustrated in Figure \ref{fig:eco}. Both systems have limitations in the type of regulations it supports. REED was build to comply with DFAR 242.204-7012 or FAR 52.204-21, and HIPAA. ResShield and ResVault were meant to be FISMA compliant. A common trend in such solutions is to use a widely accepted framework as a common language to build their ecosystems. NIST SP 800-171, the standard developed to protect CUI in nonfederal systems and organizations, becomes the best choice to support CUI.  

While the existing work laid out some foundation for campus IT to better manage CUI data, the scope and the impact of the solution are far from being sufficient. It is necessary for a broader and unified ecosystem to be developed to manage regulated data. This solution needs to be validated and tested in a few institutions before it is adopted by more colleges. The new solution should also provide cost-effective and easy-to-use CUI protection services to researchers world-class community cluster environments. 
The new solution is built on top of existing frameworks and technologies. It expands from a single computational environment to an entire ecosystem of technologies to support regulated data. As part of the success factor, undergraduate student interns from the cybersecurity or related major are recruited. These students will work side-by-side with professional IT staff and develop real-world skills applying cybersecurity to cutting-edge research cyberinfrastructure.

A single ecosystem, complete with researcher and administrator awareness programs, is needed to enable new partnerships with government agencies and industry partners.
Researchers also need a transparent framework so that they can enjoy faster intake of new funded projects; clearly understand previously complicated data security regulations; and be more competitive for research dollars. The new solution also empowers students by providing real-world implementation projects in conjunction with their classroom learning. 

\section{Users' perception on CUI}
To better capture the need for CUI training, an IRB approved survey was created to get a sense of people's perception on securing CUI. The survey followed survey design method described in \cite{weisberg1989introduction}. 
The targeted audience include undergraduate students, graduate students, faculty, research staff and IT admins. Participants were asked about their knowledge and opinions on CUI in general, and specifics of FERPA, HIPAA, NIST SP800-171, and GDPR. 
Till March 2020, we have received about 189 valid responses, with 43\% of them being faculty or research staff, and 45\% of being graduate students. Among the survey takers, only 7.4\% of the participants were confident about their knowledge in CUI. As many as 60.9\% of them admitted they know nothing about CUI or with about 20\% stated they have never even heard of CUI  before. As many as 86\% of the participants stated they don't know what laws or regulations are related to CUI, 
and 46\% of the participants were not sure if their research are related to CUI. 

For different laws and regulations, 46\% of the  participants stated they have a very good understanding of FERPA. The actual test knowledge on FERPA show 87.6\% of them does have good understandings on the basics of FERPA.  About 22\% of the population believes they know reasonable well on HIPAA. Results show 91.5\% of them can answer the technical questions on HIPAA correctly. Only 13.3\% of the participants stated they have a fair understanding of NIST SP 800-171. Knowledge tests show almost all of them correctly answered the true false question on SP 800-171. GDPR is an European Union law that does not directly apply to the U.S., unless you have cloud computing resources located in Europe. Survey showed that only 8.7\% of the participants are confident about their GDPR knowledge. Despite the fact that only 28 participants (or 14.8\%) stated that they have worked on projects related to CUI before, as many as 19.3\% of the survey takers believe share and store CUI under multiple laws and regulations are not difficult. 

We also broke the population into two groups. Group A has about 86 people and the make up of the group are faculty, research staff, and IT Admins. Group B are undergraduate students, graduate students and others. The size of Group B is 107. By comparing the answers from both groups, it shows that 67\% of the faculty group (Group A) are not confident about CUI whereas 72.6\% of the student group (Group B) claims they don't much about CUI. On the perceptions of FERPA, statistically difference between the two groups were not found. Both groups shows around 87\% of the population with a decent understanding on FERPA. Faculty group has a better knowledge in HIPAA  (18.8\%) than the student group (13.2\%). Both groups have a low understanding on NIST SP 800-171: 5.9\% for Group A and 5.6\% for Group B.  Faculty group has a slightly better knowledge on GDPR (4.7\%) than the student group (3.8\%).

From the survey, it is evident that PIs, researchers and the IT administers all lacks knowledge on CUI compliance. Since the population we surveyed are most likely being required to take FERPA training due to the mandate of university, it reveals two simple facts: 1) people can be trained to understand compliance well; and 2) training on CUI is far from being satisfactory.  The survey indicates that we must have a clear framework to help managing CUI and training must be provided and monitored to CUI stakeholders. 

\section{CUI Management Workflow}
Before introducing the CUI management workflow, it is necessary to identify the stakeholders. As stated before, CUI management is more than just an IT problem. This is evidenced by the extensive reach it involves.  

\begin{itemize}
    \item Principal Investigator (PI) and Co-Principal Investigators. PIs are ultimately responsible for the success of the program and they must make sure CUI information is compliant during the entire project life-cycle. 
    \item Sponsored Research Program/Office. Research Office will work with PIs on CUI management during the pre-award phase as well as post-award phase.   
    \item Human Research Protection Program (IRB). IRB needs to be informed to approve human subject study proposed by the PIs. Planning on CUI management should be clearly documented. 
    \item Export Controls and Research  Information Assurance (EC/IAO). Many organizations has a dedicated office to monitor and manage information assurance. 
    \item Information Technology Department Research infrastructure and data needs the support from the IT on high speed networking and high performance computing. 
    \item IT Research Computing. Some institutions have a separate research computing center to manage research data. 
    \item IT Security and Policy. CUI management eventually will be operated under the information security office of the IT department. 
    \item Platform Owner (PO). In platform terminology, the owners have control on the intellectual property and governance.
    \item Oversight Committee (OC). The OC usually serves as the highest-level decision-making group on information security issues at the University. 
\end{itemize}

\begin{figure}[tb]
    \centering
    \includegraphics[width=.5\textwidth]{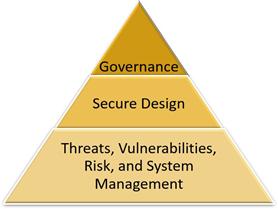}
    \caption{Realms}
    \label{fig:Realms}
\end{figure}

The CUI Framework is broken into three phases, or realms. They are 1) governance, 2) secure design, and 3) threats, vulnerabilities, risk, and system management, as shown in Figure \ref{fig:Realms}. A total of 11 steps are further defined under the three realms. \\

\textbf{Phase 1. Governance, Risk, and Compliance, Security Awareness Alignment} \\
There are four steps in phase 1.  \\
\textit{Step 1. Policy and Governance}. Major tasks in step 1 are as follows: 
\begin{itemize}
    \item Define the scope for your program. Create or adopt a change management process.
    \item Define Roles. Assign primary roles, secondary/oversight roles.
    \item Perform policy reviews.
    \item Create a unified glossary.
    \item Create unified templates. 
\end{itemize}

\textit{Step 2. Review Current Business Processes}. The following will be reviewed in step 2:
\begin{itemize}
    \item Workflows.
    \item Staffing levels.
    \item Additional Requirements.
    \item Training Requirements. 
\end{itemize}

\begin{figure}[tb]
    \centering
    \includegraphics[width=0.98\textwidth]{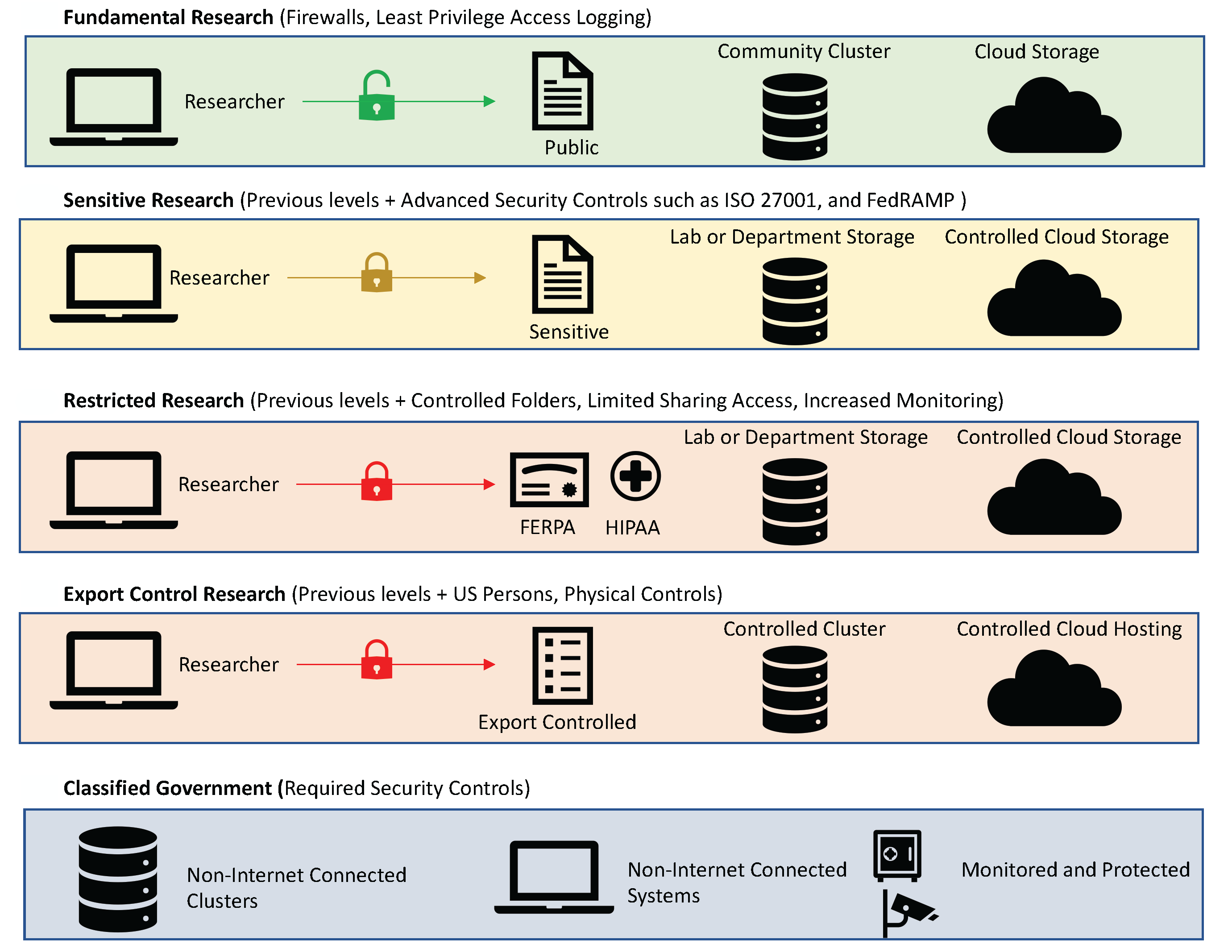}
    \caption{Research Data Security}
    \label{fig:RdataSec}
\end{figure}

\textit{Step 3. Create Security Controls}. Major tasks in step 3 are listed below:
\begin{itemize}
    \item Classify data security levels based on regulator requirement, details shown in Figure \ref{fig:RdataSec}.
    \begin{itemize}
        \item Fundamental Research
        \item Sensitive Research
        \item Restricted Research (HIPAA and FERPA)
        \item Export Controlled Research 
        \item Classified government
    \end{itemize}
    \item Create a process to ensure controls meeting the threat level of existing and newly discovered vulnerabilities; generate threat model for research. 
    \item Create Placeholder for SOC requirements.
\end{itemize}

\textit{Step 4. Create Training Program}. Step 4 involves the following tasks:
\begin{itemize}
    \item Review modify current training options if required.
    \item Reviews current trends and immediate risks.
    \item Review compliance training requirements.
    \item Determine how training will be offered.
    \item Create any required training materials. 
\end{itemize}

\vspace{6pt}
\hspace{-12pt} 
\textbf{Phase 2. Alignment Process, Business Continuity and Disaster Recovery} \\
Step 5 - 9 are defined in phase 2.  \\
\hspace{-12pt} 
\textit{Step 5. Categorize Information Systems}. Major tasks in step 5 are as follows: 
\begin{itemize}
    \item Categorize information system, such as type of data and operational requirements. 
    \item Described the information system.
    \item Start drafting security plans.
    \item Start risk assents for new systems.
    \item Review risk assessments for existing systems.
    \item Added to asset inventory.
\end{itemize}

\begin{figure}[bt]
\centering
\includegraphics[width=1.05\textwidth]{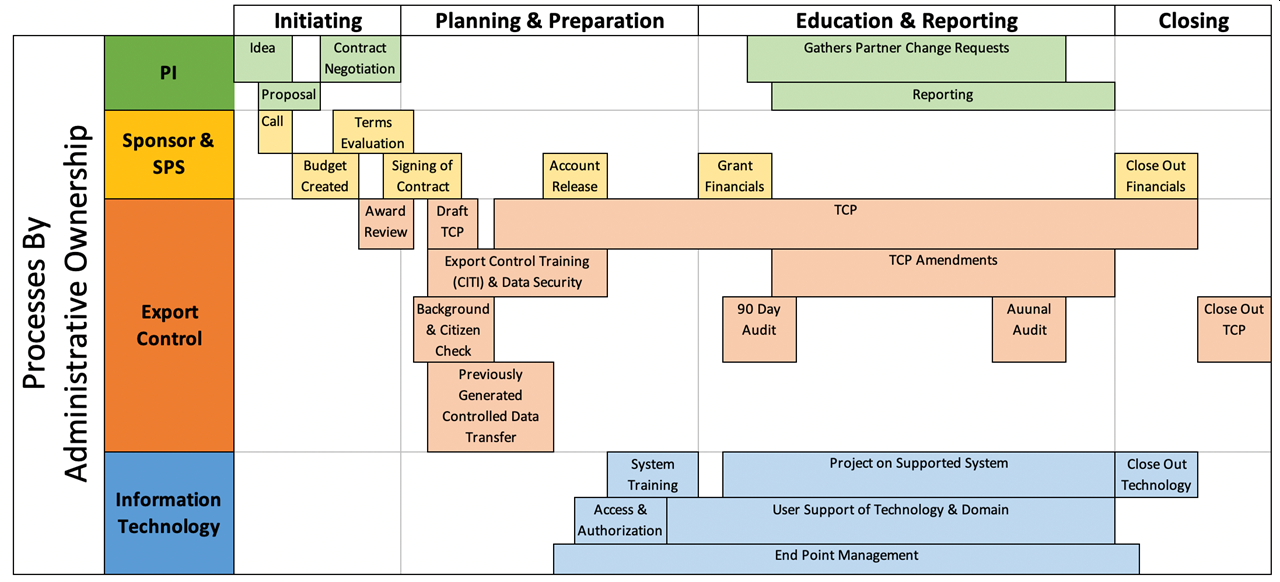}
\caption{Single Take Process Map}
\label{fig:process}
\end{figure}

\textit{Step 6. Select the Defined Security Controls}. Step 6 involves the following tasks: 
\begin{itemize}
    \item Determine access level.
    \item Tailor controls if required.
    \item Add to monitoring strategy.
    \item Complete new Risk Assessments. 
    \item Complete Security Plans
\end{itemize}

\textit{Step 7. Implement Security Controls}. Major tasks in step 7 are listed below:  
\begin{itemize}
    \item Implement security controls.
    \item Amend security plan if required. 
    \item Communicate completion date to stakeholders. 
    \item Complete required compliance training. 
\end{itemize}

\textit{Step 8. Access Security Controls}. Step 8 involves the following tasks: 
\begin{itemize}
    \item Test the security controls prior to go live. 
    \item Create a security report outlining any findings.
    \item Adjust any findings, or determine its safe to operate.
    \item Add the remaining findings to risk assessment or adjust the security plan.
\end{itemize}

\textit{Step 9. Authorize Information System}. Specific tasks in this step include the following: 
\begin{itemize}
    \item Document residual risk using a Plan of action and milestones (POAM) template. 
    \item Collect all reports, documentation and store it in a secure location for future reference.
    \item Complete risk acceptance and sign the system security plan and distribute.
    \item Work with Regulator Offices and Contracts, providing the required details to recommend or authorize the use of the service. 
\end{itemize}

\vspace{24pt}
\textbf{Phase 3. Threat Intelligence, Audits/Alerts, Incident Management and Response, Security Awareness, Communications, Security Training, Operations, Professional and Organizational Development} \\
Two more steps are defined in phase 3.  \\
\textit{Step 10. Monitor Security Controls}. Monitoring is a comprehensive step that includes, and not limited to, the following tasks: 
\begin{itemize}
    \item Monitor system and environment changes.
    \begin{itemize}
        \item change management
        \item vulnerability scanning
        \item vendor notification
        \item port scanning
        \item log review
    \end{itemize}
    \item Review access controls.
    \item Mediate findings.
    \item Update required documentation, such as System Security Plan and POAM.
    \item Decommission
    \item Incident Response
\end{itemize}

\textit{Step 11. Conduct and Access Training}. This step includes the following tasks:
\begin{itemize}
    \item Provide conduct training both in-person and on-line. 
    \item Conduct communications based on immediate risk.
    \item Send notification to individuals missing mandatory training. 
    \item Conduct Table Top Exercise with new Monitoring tools and incident response. 
    \item Measure training results.
\end{itemize}

When working through our Institution's "Step 2: Review current workflows", Figure \ref{fig:process} resulted demonstrating the end-to-end workflow for a Researcher with CUI contracts. Each process within this entire workflow is supplemented with documentation and support.


\section{Beyond the Framework}
While the outcome of a CUI framework is to help an institution meet the compliance requirement, the impact goes beyond the framework itself. 

\vspace{6pt}
\hspace{-12pt}
\textbf{Build a Cybersecurity Culture}\\
Security awareness training is created specifically for researchers and professional development for engineers, analysts, and research administration staff. This brings in a whole new group of people and stakeholders and we want to ensure that this does not cause extra work for the scientist or scholar or take away time from what they do best. Clear business processes and their documentation, as well as training, are developed to ensure that everyone knows what they are supposed to do and where their work fits into the whole of the research data life cycle. 
Many of the challenges are not technical. They deal with measuring risk, and setting priorities. Institutions need to  provide an open communication model, and a central source for compliance and security questions. Security by design is the concept should be embraced at the institutional level, and  adopted by the individual researchers. Key components required to build such a successful cybersecurity culture are as follows. 

\begin{itemize}
    \item NIST-based framework with defined security controls
    \item Threat model specific for research ecosystems
    \item Purpose built platforms for regulated ecosystem including 
    Secure High Performance Computing (HPC)
    \item Common templates and guidelines based on CTSC Cybersecurity Program Guide
    \item Culture driving security awareness program
    \item Self-managed platforms subject to institutional policies and control
    \item Clear ownership and responsibility of all aspects of data security for data housed and processed within this environment
    \item Proper Full Time Equivalent (FTE) count and automation
    \item Sufficient secure storage, both active and archival storage
    \item Unified Funding Model
 
\end{itemize}

Periodically review and assess key compliance metrics is an important aspect of the cybersecurity culture. 
The most recent internal annual check-up showed that 100\% of HIPAA, GLBA, FERPA reviews were performed within eight weeks; 100\% of Technology Control Plan were reviewed and approved within 5 business days; 100\% of Security reviews were completed within 30 Days; and 78\% of Security Exceptions are Current, an improvement from 76\% in year 2020.

Additional functionality is expected to supplement the established framework over time:
\begin{enumerate}
   \item The ability to provide a common security culture to all IT staff in Schools and Colleges
    \item Automation of common tasks
    \item Create a maturity score for a platform to help define the state of security control adoption
    \item Standardized way to categorize and describe systems
    \item Unified data management plans
    \item Build trust and expertise within the University and with students
\end{enumerate}

\textbf{Grow the Community}  \\
To share our work with institutions facing similar challenges, We hosted six workshops and recruited a group of research intense institutions within the U.S. 
Due to Covid-19, The workshops were arranged 
with facilitated services, which allowed the group to easily perform dynamic brainstorming and generated a full repository of discussions. 
In total, there were 155 participants from 84 different institutions attended at least one of workshops. The workshops were scheduled every six to eight weeks, with each workshop lasting for about 2 hours. Participants discussed the challenges, gaps, possible actions, best practices. Feedback was collected and the summary of the workshop was documented in the community report \cite{cuiworkshop2021}.
This consensus based white paper became a valuable interdisciplinary product that involves applied technology, policy making, and legal compliance.
The workshop series on CUI greatly enhanced the awareness of the community on the subject. The post workshop survey showed 76\% of the 50 responses found the
workshop ‘very useful’; 24\% found it to be ‘useful’; 0\% stated it was ‘not very useful’. In addition, 89.6\% of the 48 responses believed that the virtual event influenced the way how they approach compliance at their institutions. More excitingly, 90.9\% of the 44 responses stated the virtual events introduced them to new collaboration opportunities. With the dominating positive feedback from the community, we have a strong reason to believe the improved practice derived from this project will help other institutions to adopt and revise their workflow to stay effective. 

\hspace{-12pt}
\textbf{Cybersecurity Training and Student Engagement}\\
The cybersecurity field is currently facing a severe workforce shortage. According to Herjavce Group \cite{ventures2017cybersecurity}, there will be  more than 3.5 million unfilled cybersecurity job openings by 2021, and cybercrime will be more than three times the number of job openings over the next 5 years. The CUI framework project aims to grow the cybersecurity workforce by recruiting and training undergraduate students in the deployment and operation of networking hardware and software. Eight undergraduate students and three graduate students have worked on this project. The undergraduate students were recruited from the cybersecurity or closely related major. Given that only 11 percent of the cybersecurity workforce is comprised of women \cite{pusey2016outcomes}, the recruitment process is therefore targeting those underrepresented groups the cybersecurity field, especially women and minorities.

Training and educational materials are developed by following the recommendations for CUI Basic Training from the Information Security Oversight Office \cite{cuiTraining2018}.
All materials are be designed with clearly identified learning objectives, using Bloom’s Taxonomy \cite{bloom2009taxonomy}. Materials address the stated learning objective and assessments devised to check for learner ability to meet the stated learning objectives. Our next step is to craft pre-lesson and post-lesson assessments and delivered in order to determine increase in knowledge as a result of lesson materials.

In order to increase the efficacy of the education effort, materials are delivered in a multi-modal format to suit multiple cognitive styles \cite{sankey2010engaging}. In-person workshop, online text based as well as online video based lessons are developed and delivered. Recognizing that training staff is only a part of the necessary training and education required to create a more secure institution, a security awareness program is currently under development. This will ensure that awareness and education about secure handling of controlled unclassified information will be part of the culture and conversation year-round versus only during annual certification periods. The design and delivery of the security awareness follows the recommendations described in \cite{kim2014recommendations}.
Student interns play a vital role in the development, delivery and dissemination of educational materials and the security awareness program.


\section{Analysis on Project Contract Approval Process}
After four years being deployed on Purdue campus since 2018, the framework has successfully and effectively helped Purdue staffs to review and complete over thousands of federal project contracts in a timely manner. According to the Single Take Process Map in Figure 4, project contracts need to be signed by Sponsored Research Program/Office. Depending on the project security level, some contracts could be required for review by Export Control. Typically it takes at least 14 days for Sponsored Research Office to process the proposal of contract, and project contracts with Export Control require extra 30-90 days in the approval process than those without Export Control. In Figure 5, the data of Federal contracts from Purdue University collected in last four years shows that using the framework, average 75\% of total Federal contracts per year (including contracts with and without Export Control) were completed within 30 days. And 14\% of contracts approved within 30 days were reviewed by Export Control, and the rest of contracts weren't. 
\begin{figure}
\centering
\includegraphics[width=1\textwidth]{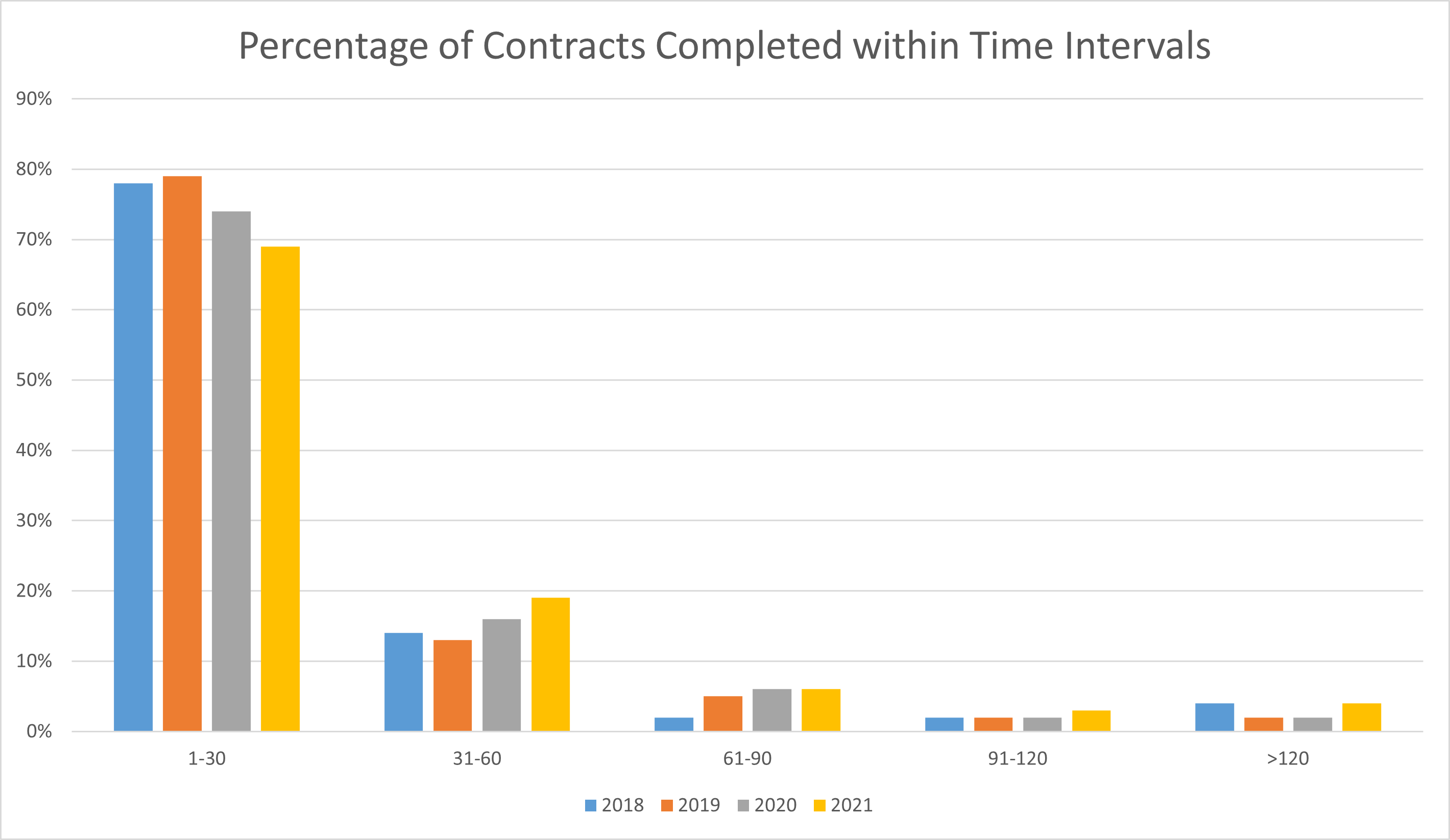}
\caption{Percentage of Contracts Completed within Time Intervals}
\label{fig:percent}
\end{figure}
In Figure 6, the average days of completing contracts with or without Export Control were presented. The average days of completing a contract without Export Control for last four years was 22.28, and the average days of completing a contract with Export Control was 39.24. As the framework has been applied on Purdue campus for four years, the contract approval process has achieved faster speed and promoted more effective collaboration between different departments for less delay.
\begin{figure}
\centering
\includegraphics[width=1\textwidth]{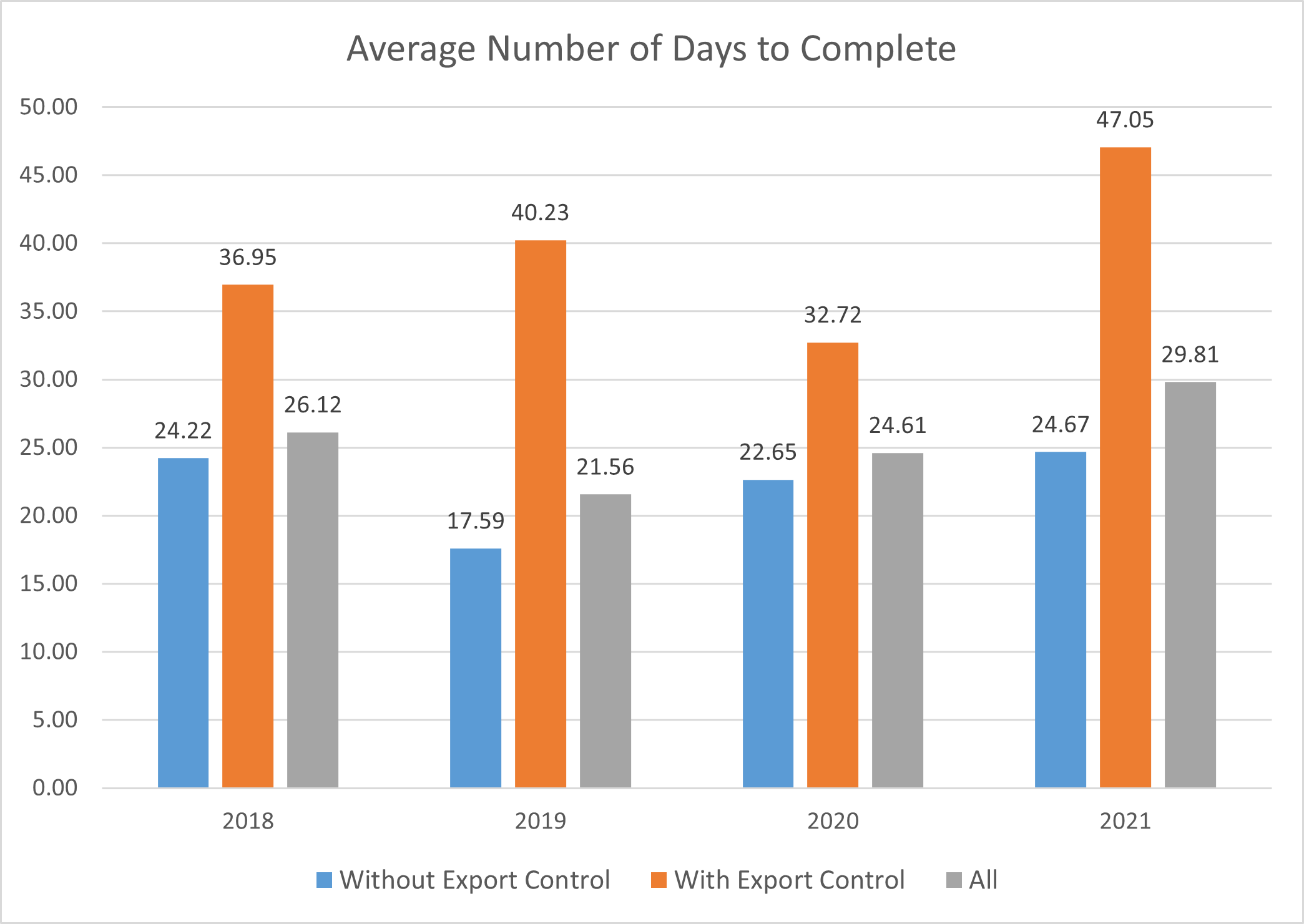}
\caption{Average Days to Complete Contracts}
\label{fig:avgComp}
\end{figure}

\section{Conclusions and Future Work}
Managing CUI is certainly not an easy task. It crosses multiple sectors in a campus environment and needs the support from researchers, IT Admins, staff in Sponsor Program office, officers in compliance office, and the administrators of the institution. The workflow presented in this work demonstrated a scalable solution that be applied to most research institutions in the U.S. Because fundamental similarities exist between NIST SP 800-171 and ISO 27001/27002, this work can be further extended to the research institutions outside the U.S.. In addition to the framework, baselines for understanding and awareness are established. Annual assessments of key staff and research administrative staff are conducted. 

Once an institution has an established cybersecurity framework, it can have the ability to measure success using its data analytics, logging, and incidents as key performance indicators. These data points will help the leadership provide an ongoing strategy for security posture. The various working groups can use surveys, town hall meetings, and information sharing to measure success. This will help them to outline new controls and training models.

Managing CUI is a moving target. The framework illustrated in this work will be periodically reviewed and revised so that it meets the highest standard in cybersecurity and is lean and mean from management perspective. In the future, we will disseminate more matured version of this framework, along with the awareness program and other training materials to other institutions. Our goals is that other colleges do not have to rebuild the wheels. Instead, they can build their success on top of this framework. Together, we will form a bigger collation and ecosystem to manage sensitive research data under a common scheme.


\section*{Acknowledgments}
This work is sponsored by 
Nation Science Foundation (NSF), award number 1840043. 
The authors would like to thank undergraduate students for their contributions to this work. 

\clearpage
\bibliographystyle{unsrtnat}  
\bibliography{Arxiv-CUI}

\end{document}